\documentclass[
reprint,
superscriptaddress,
amsmath,
amssymb,
aps,
pra, 
longbibliography,
floatfix
]{revtex4-1}
\usepackage{hhline}
\usepackage{graphicx}
\usepackage{xcolor}

\usepackage{tabularx}
\usepackage{dcolumn} 
\newcolumntype{d}[1]{D{.}{.}{#1}}

\usepackage{booktabs}

\usepackage{comment}
\usepackage{enumitem}

\begin{document}

\title{Analogy between the magnetic dipole moment at the surface of a magnetoelectric and the electric charge at the surface of a ferroelectric.}

\author{Nicola~A.\ Spaldin}
\affiliation{Department of Materials, ETH Zurich, CH-8093 Z\"{u}rich, Switzerland}

\date {\today}

\begin{abstract}
In honor of Igor Dzyaloshinskii on his 90th birthday, we revisit his pioneering work on the linear magnetoelectric effect in light of the modern theory of ferroelectric polarization. We show that the surface magnetic dipole moment associated with magnetoelectric materials is analogous to the bound surface charge in  ferroelectrics, in that it can be conveniently described in terms of a bulk magnetoelectric multipolization that is analogous to the ferroelectric polarization. We define the intrinsic surface magnetization to be this surface magnetic dipole moment per unit area, and provide a convenient recipe for extracting it for any surface plane, from knowledge of the bulk magnetic order. We demonstrate the procedure for the prototypical magnetoelectric material, Cr$_2$O$_3$, in which Dzyaloshinskii first identified the linear magnetoelectric effect, and compare the value of the intrinsic surface magnetization to recent experimental measurements. Finally, we argue that non-magnetoelectric antiferromagnets whose multipolization lattices do not contain zero should have an intrinsic surface magnetization, in the same way that non-polar insulators whose polarization lattices do not contain zero have an intrinsic surface charge.   
\end{abstract}

\maketitle

\section{Introduction}

In a linear magnetoelectric material, an applied electric field induces a magnetization linearly proportional to the field strength, and an applied magnetic field induces a corresponding linear electric polarization. The first mention of the phenomenon, to our knowledge, is in the original 1958 edition of the classic {\it Electrodynamics of continuous media} by Landau and Lifshitz \cite{Landau/Lifshitz:1958}, with the brief statement that an effect resulting from a linear relation between the magnetic and electric fields in a substance is possible in principle. Soon after, Dzyaloshinskii proved using symmetry arguments that the behavior should occur in chromia, Cr$_2$O$_3$
\cite{Dzyaloshinskii:1960}. This was then the material of choice for the first experimental demonstration of the linear magnetoelectric effect by Astrov \cite{Astrov:1960}.

A symmetry requirement for the existence of a linear magnetoelectric response is that both time-reversal, $T$, and space-inversion, $P$, symmetries are broken. This condition is the same as that for a non-zero magnetoelectric multipole tensor, ${\cal M}_{ij} = \int r_{i} {\mu_j(\vec{r})}d^3 \vec{r}$, which is the second order coefficient in the multipole expansion of the energy of a spatially varying magnetization, $\vec{\mu}(\vec{r})$, in a spatially varying magnetic field, $\vec{H}(\vec{r})$ \cite{Ederer/Spaldin:2007,Spaldin/Fiebig/Mostovoy:2008,Spaldin_et_al:2013}: 
\begin{eqnarray}
{\rm Energy} & = & -  \int \vec{\mu}(\vec{r}) \cdot
\vec{H}(\vec{r}) \, d^3 \vec{r} \\
 & = & - \int \vec{\mu}(\vec{r}) \cdot \vec{H}(0) \, d^3 \vec{r} \label{Eqn:dipole}\\
 &  & 
-  \int r_{i} {\mu}_{j} (\vec{r}) \partial_{i} H_{j}(0) \, d^3 \vec{r} \quad  - \ldots \nonumber
\end{eqnarray}
Here, the expansion in powers of the field gradients is calculated at some arbitrary reference point $\vec{r}
= 0$, and $i,j$ are Cartesian directions with summation over repeated indices implied. The usual magnetic dipole moment, $\vec{m} = \int \vec{\mu}(\vec{r}) d^3 \vec{r}$ appears in the first term of the expansion of Eqn.~\ref{Eqn:dipole}. The ${\cal M}_{ij}$ tensor in the second term is often discussed in terms of its irreducible components, the scalar magnetoelectric monopole, 
\[a =  \frac{1}{3} {\cal M}_{ii} = \frac{1}{3} \int \vec{r} \! \cdot
\vec{\mu}(\vec{r}) d^3 \vec{r} \quad , \] 
the toroidal moment vector, 
\[\vec{t} = \frac{1}{2}  \int \vec{r} \! \times \vec{\mu}(\vec{r}) d^3 \vec{r}\] 
with components 
\[t_i = \frac{1}{2} \varepsilon_{ijk} {\cal M}_{jk} \quad , \] 
and the traceless quadrupole tensor, 
\begin{eqnarray}
q_{ij} &=& \frac{1}{2}\left({\cal M}_{ij} + {\cal M}_{ji} - \frac{2}{3} \delta_{ij} {\cal
M}_{kk}\right)\nonumber\\
&= &\frac{1}{2} \int \left[r_i \mu_j +
r_j \mu_i - \frac{2}{3} \delta_{ij} \vec{r}\! \cdot \vec{\mu}(\vec{r})
\right] d^3\vec{r} \quad ,
\end{eqnarray} 
such that
${\cal M} = $
\begin{equation}
\begin{bmatrix}
a + \frac{1}{2}q_{x^2-y^2} - \frac{1}{2} q_{z^2} 
& t_z + q_{xy} & t_y + q_{xz}\\
-t_z + q_{xy} & a - \frac{1}{2}q_{x^2-y^2} - \frac{1}{2} q_{z^2}  & -t_x + q_{yz} \\
-t_y + q_{xz} & t_x + q_{yz} &  a + q_{z^2} 
\end{bmatrix}
.
\end{equation}

\begin{figure}
\centering
\includegraphics[scale=0.3]{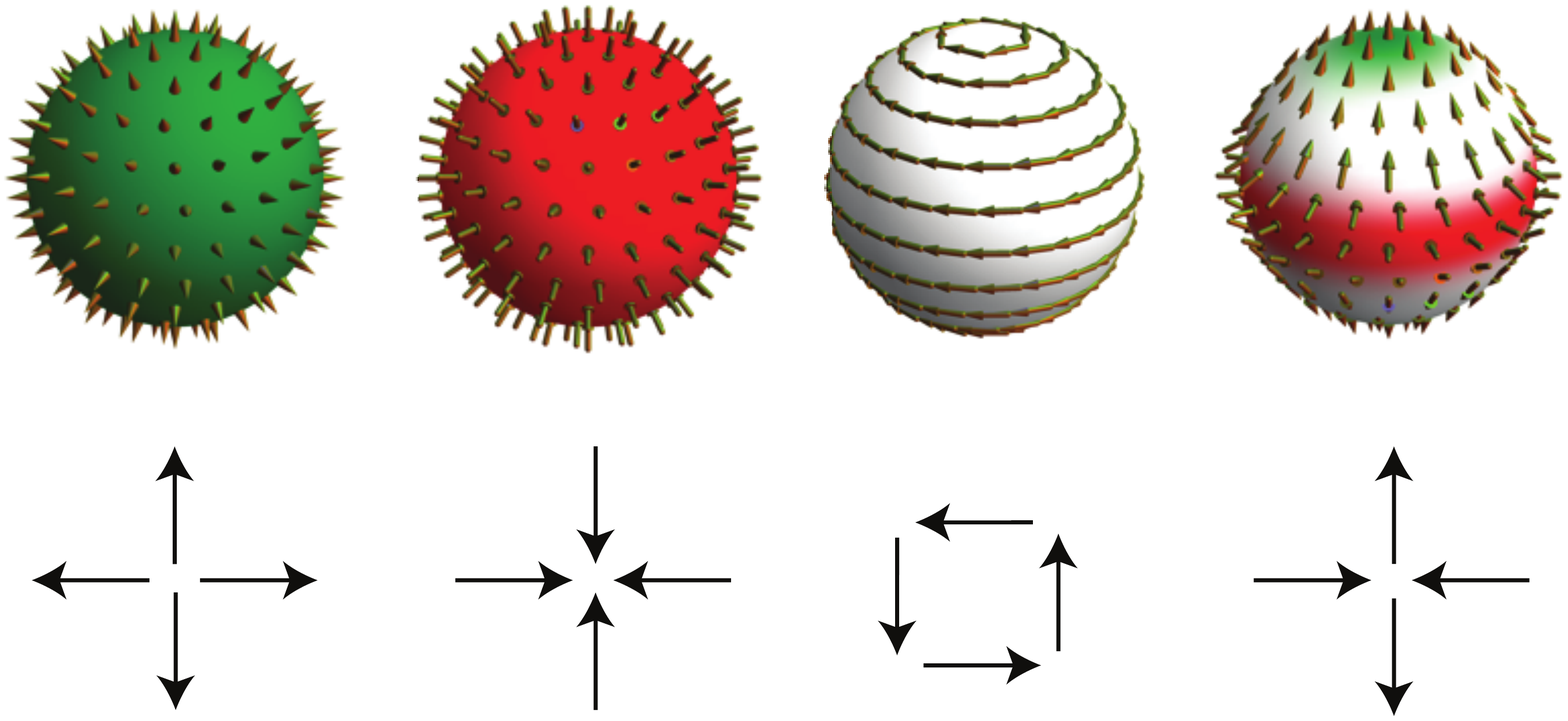}
\caption{Cartoons of (left to right) positive and negative magnetoelectric monopoles, the $z$ component of the toroidal
moment and the $z^2$ component of the quadrupole moment. The upper panel (reproduced from Ref.~\onlinecite{Spaldin_et_al:2013}) shows magnetoelectric multipoles formed in the magnetic texture in the sphere around an atom or ion, with the small gold arrows indicating the orientation of the magnetization at each position. The black arrows in the lower panel indicate local magnetic dipole moments that combine at the unit-cell level to form magnetoelectric multipoles. These two cases were called the atomic-site and local-moment contributions in Ref.~\onlinecite{Spaldin_et_al:2013}.
\label{Hedgehogs}}
\end{figure}

Representative monopoles, toroidal moments and quadrupoles are sketched schematically in Fig.~\ref{Hedgehogs}. Note that these non-centrosymmetric magnetization textures can form within the charge cloud around an ion (upper panel of Fig.~\ref{Hedgehogs}) or as a result of the distribution of magnetic dipoles within the system (lower panel of Fig.~\ref{Hedgehogs}). We will focus on the latter contribution here, since it has been shown to be the dominant contribution in typical transition-metal oxides \cite{Spaldin_et_al:2013}.

When appropriately normalized by the volume in the case of bulk, periodic systems we call these irreducible components the monopolization, $A = \frac{a}{V}$, toroidization, $\vec{T} = \frac{\vec{t}}{V}$, and quadrupolization, $Q_{ij} = \frac{q_{ij}}{V}$, respectively, by analogy with the magnetization or ferroelectric polarization; when we refer to them collectively we will use the term {\it magnetoelectric multipolization}. They provide a bulk, thermodynamic quantity associated with ``magnetoelectricness'', complementing the usual definition of magnetoelectricity as a response function. 

One property in which this bulk thermodynamic magnetoelectric aspect manifests, is in the preferred orientation of the magnetoelectric multipole in combined electric and magnetic fields, described by a term in the free energy of the form
\begin{equation}
F_{\text{ME}} \propto - {\cal M}_{ij} E_i H_j \quad .
\end{equation}
This preferred-orientation property can be exploited to prepare magnetoelectric antiferromagnets in single-domain states by annealing them in combined electric and magnetic fields, a process known as magnetoelectric annealing \cite{Shtrikman/Treves:1963}. For example a magnetoelectric monopole of positive (negative) sign is favored by parallel (anti-parallel) electric and magnetic fields, and a toroidal moment will preferentially align along the cross product vector of $E$ and $H$ fields. The emerging field of antiferromagnetic spintronics uses such combined magnetic and electric fields to lift the degeneracy of otherwise equivalent antiferromagnetic domain states and prepare samples with well-defined single antiferromagnetic domains of selected orientation \cite{Binek/Doudin:2005,Borisov_et_al:2005}.

A second scenario in which this thermodynamic aspect manifests, which was pointed out by Dzyaloshinskii in 1992 \cite{Dzyaloshinskii:1992}, is in the power-law decay of the external magnetic field around an antiferromagnetic material with a net non-zero magnetoelectric multipolization. Power-law behavior is fundamentally different from the exponential field decay expected around a conventional centro- or time-reversal symmetric antiferromagnet \cite{Dzyaloshinskii:1992}. In the particular case of the prototypical magnetoelectric Cr$_2$O$_3$, which has non-zero magnetoelectric monopolization and quadrupolization below its N\'eel temperature, Dzyaloshinskii showed that the external field should have the angular form of a magnetic quadrupole. As in the case of the original magnetoelectric response prediction, this was subsequently confirmed by Astrov \cite{Astrov/Ermakov:1994,Astrov_et_al:1996}, although the measured field strength was smaller in magnitude than predicted. 
The intrinsic bulk nature of the measured external field dependence was subsequently questioned, however, when it was pointed out that any antiferromagnet can in principle have a {\it surface} magnetization that, depending on the sample shape and choice of surface termination, could give rise to a magnetic field \cite{Andreev:1996}. The discussion was further enriched by recent theoretical demonstrations that certain surfaces of a magnetoelectric antiferromagnet will {\it always} have a surface magnetization \cite{Belashchenko:2010} and associated external magnetic field \cite{Jiang/West/Zhang:2020} as a consequence of the bulk magnetoelectric multipolization.

The discussion of whether the external magnetic field associated with a finite-sized sample of magnetoelectric material is a surface or a bulk property might at first sight seem rather academic. Indeed, from a fundamental point of view, the situation in magnetoelectric materials today is reminiscent of that in the field of ferroelectric materials fifty years ago \cite{Martin_comment:1972}, before the modern theory of polarization was developed \cite{Resta:1993,King-Smith/Vanderbilt:1993}, and the bulk nature of the ferroelectric polarization was clearly established. In particular, a universally accepted definition of surface or boundary magnetization is still lacking. In addition to its fundamental interest, the question of whether antiferromagnets possess a surface magnetization, how it should be defined, and whether it is fundamentally different in magnetoelectric and non-magnetoelectric antiferromagnets,  is also potentially technologically relevant. Antiferromagnets are widely used in magnetic sensors and data storage devices to pin the orientation of adjacent ferromagnetic layers so that they are not readily reoriented by  magnetic fields. Many details of this exchange-bias phenomenon remain to be understood \cite{Nogues/Schuller:1999}, and recognizing  a fundamentally different behavior between the surfaces of magnetoelectric and non-magnetoelectric antiferromagnets might contribute to this understanding. 

In this paper, we approach the description of the surface magnetism of magnetoelectric antiferromagnets by making a correspondence with the surfaces of ferroelectrics. We argue that, just as the surface of a ferroelectric carries a well-defined bound charge resulting from its bulk polarization \cite{Vanderbilt/King-Smith:1993}, the surface of a magnetoelectric carries a magnetic dipole moment resulting from its bulk magnetoelectric multipolization. We define the {\it intrinsic surface magnetization} to be this surface magnetic dipole per unit area. We emphasize that it arises from a bulk property, the magnetoelectric monopolization, and that it can be uniquely defined for a particular choice of surface plane orientation and termination.  We discuss the analogy between the polarization quantum and polarization lattice, which are consequences of the periodicity in bulk ferroelectrics, and the analogous properties in periodic bulk magnetoelectrics, which we call the magnetoelectric multipolization increment and magnetoelectric multipolization lattice. Finally, we show that the description is also relevant for {\it non-magnetoelectric} antiferromagnets, allowing a classification into one of two types with fundamentally different surface magnetic properties: the trivial case, in which the lattice of magnetoelectric multipolization values contains zero, and non-trivial antiferromagnets whose magnetoelectric multipolization lattices do not contain zero, in spite of their not being magnetoelectric. 

\section{The surface charge arising from ferroelectric polarization and the surface magnetic dipole moment arising from magnetoelectric multipolization.}

\begin{figure}
\centering
\includegraphics[scale=0.45]{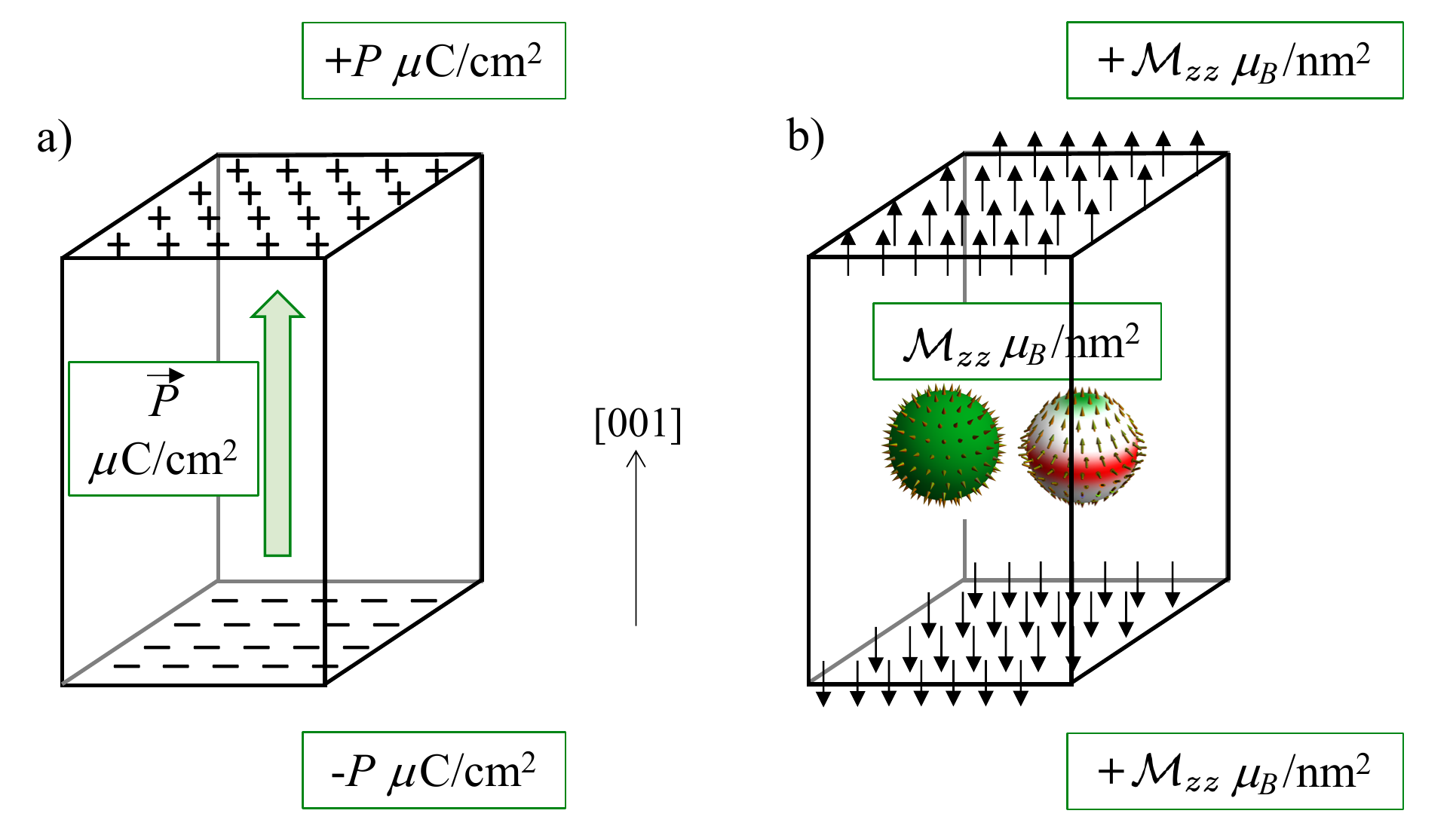}
\caption{a) Surface charge associated with ferroelectric polarization, $\vec{P}$. b) Surface magnetic dipole moment associated with magnetoelectric multipolization, ${\cal{M}}_{zz}$. The $-$ signs, $+$ signs and small black arrows on the surfaces indicate negative charge, positive charge and magnetic dipole moments. The ferroelectric has negative charge on its lower surface and positive charge on its upper surface; the magnetoelectric has positive magnetic dipole moments (pointing outwards from the sample) on both its upper and lower surfaces.}
\label{P_vs_Mzz}
\end{figure}

From a quick glance at the units of electric polarization, which are dipole moment per unit volume or equivalently charge per unit area, it is clear that a surface perpendicular to the polarization direction in a ferroelectric material carries a bound charge per unit area equal to the value of the polarization, with the sign of the surface charge given by the direction of polarization, as shown in Fig.~\ref{P_vs_Mzz}a. (For a rigorous derivation see Ref.~\onlinecite{King-Smith/Vanderbilt:1993}). 
This surface charge is typically screened by the metallic electrodes in a capacitor geometry, or by surface adsorbates on a free surface, where it gives rise to the pyroelectric effect. The surface charge can be quite large -- in the prototypical example of the II-IV perovskite-structure PbTiO$_3$, in which the polarization is along the [001] axis, it has the value $\sim$60 $\mu$C/cm$^{2}$, corresponding to slightly more than half an electronic charge per unit cell, on the (001) surface -- and inadequate screening in a thin-film geometry gives rise to a depolarizing field which drives domain formation \cite{Fong_et_al:2004}, reorientation of the polarization direction into the plane, or can even stabilize exotic skyrmion-like states \cite{Das_et_al:2019}. 
We note that this surface charge can not be compensated by reconstructions of the surface without the addition or removal of charged species.

While the ferroelectric polarization has units of charge per unit area, the magnetoelectric multipolization, or magnetoelectric multipole per unit volume, has units of magnetic dipole moment per unit area. Therefore, by analogy with the ferroelectric case, the surface of a magnetoelectric should have a magnetic dipole moment per unit area, whose size and orientation depends on the {\it bulk} magnetoelectric multipolization. We refer to this as the intrinsic surface magnetization, since it results from a bulk property of the material; 
it is this surface magnetization that was discussed in Ref.~\onlinecite{Belashchenko:2010}. In Fig.~\ref{P_vs_Mzz}b we illustrate the analogy with ferroelectricity for the case of a uniaxial magnetoelectric with non-zero monopolization and  $z^2$ quadrupolization, whose in-plane contributions cancel so that the only non-zero component of the magnetoelectric multipolization tensor is ${\cal{M}}_{zz}$. (We will see later that this is the case in Cr$_2$O$_3$). The ${\cal{M}}_{zz}$ component results in a $z$-oriented magnetic moment on those surfaces whose surface normal has a component along the $z$ axis, that is the $(001)$ and $(00\bar{1})$ surfaces in this example. A positive ${\cal{M}}_{zz}$ indicates that the moments on surfaces with a normal component along $+\vec{z}$ are oriented along $+\vec{z}$, and those on surfaces with a normal component along $-\vec{z}$ are oriented along -$\vec{z}$, and vice versa. We note that, just as the ferroelectric case in (a) has no net charge, the net magnetization of the magnetoelectric in (b) is zero.

In Table~\ref{E_vs_M_Multipole} we list the first three terms of the multipole expansion, discussed above for the case of the magnetic field, for the electric-field and magnetic-field cases. To facilitate the comparison between the two cases, we write the magnetic expansion coefficients in terms of their current, rather than their magnetization, densities. The purpose of the Table is to emphasize that, while the bulk property that gives rise to a surface charge (the zeroth-order term in the multipole expansion) is the  dipole (the first-order term in the multipole expansion), the bulk property that gives rise to a surface dipole is the {\it second-order} term in the multipole expansion. In this context, the surface magnetic dipole does not arise from the bulk magnetization (which is the magnetic dipole moment per unit volume), but from the bulk magnetoelectric multipolization. Conversely, the bulk magnetoelectric multipolization can be used to provide an unambiguous definition of the surface magnetic dipole moment.

\begin{table*}
\renewcommand{\arraystretch}{1.4}
\setlength{\tabcolsep}{10pt}
\begin{tabular}{c|ccc} \hline \hline
    & 2nd order & 1st order & 0th order  \\ \hline 
    &  electric quadrupole     & electric dipole   &  electric charge         \\
$\vec{E}$ &  $q_{ij}=\int r_ir_j\rho(\vec{r}) d^3 \vec{r}$           & $\vec{p} = \int \vec{r} \rho(\vec{r}) d^3 \vec{r}$           & $q_e = \int \rho(\vec{r}) d^3 \vec{r}$  \\ 
 & & $\rightarrow$ surface charge & \\ \hline
  
    &  magnetoelectric multipole & magnetic dipole         & magnetic monopole  \\
$\vec{H}$ &  ${\cal{M}}_{ij} = \int r_i (\vec{r} \times \vec{J}(\vec{r}))_j  d^3 \vec{r} $           & $\vec{m} = \int \vec{r} \times \vec{J}(\vec{r}) d^3 \vec{r}$           & $q_m = \int \vec{J}(\vec{r}) d^3 \vec{r}$              \\
 & $\rightarrow$ surface dipole & & \\\hline 

  \end{tabular}
\caption{Components of the multipole expansion of a general charge distribution, $\rho(\vec{r})$, in an electric field, $\vec{E}$ (top row) and of a general magnetic distribution in a magnetic field, $\vec{H}$ (bottom row). To facilitate comparison, the magnetic terms are written in terms of their corresponding current densities, $\vec{J}(\vec{r})$. We see that an $n$th order multipole has the corresponding $(n-1)$th property associated with its surface.}
\label{E_vs_M_Multipole}
\end{table*}

\section{The polarization quantum and the magnetoelectric multipolization increment}

\begin{figure}
\centering
\includegraphics[scale=0.3]{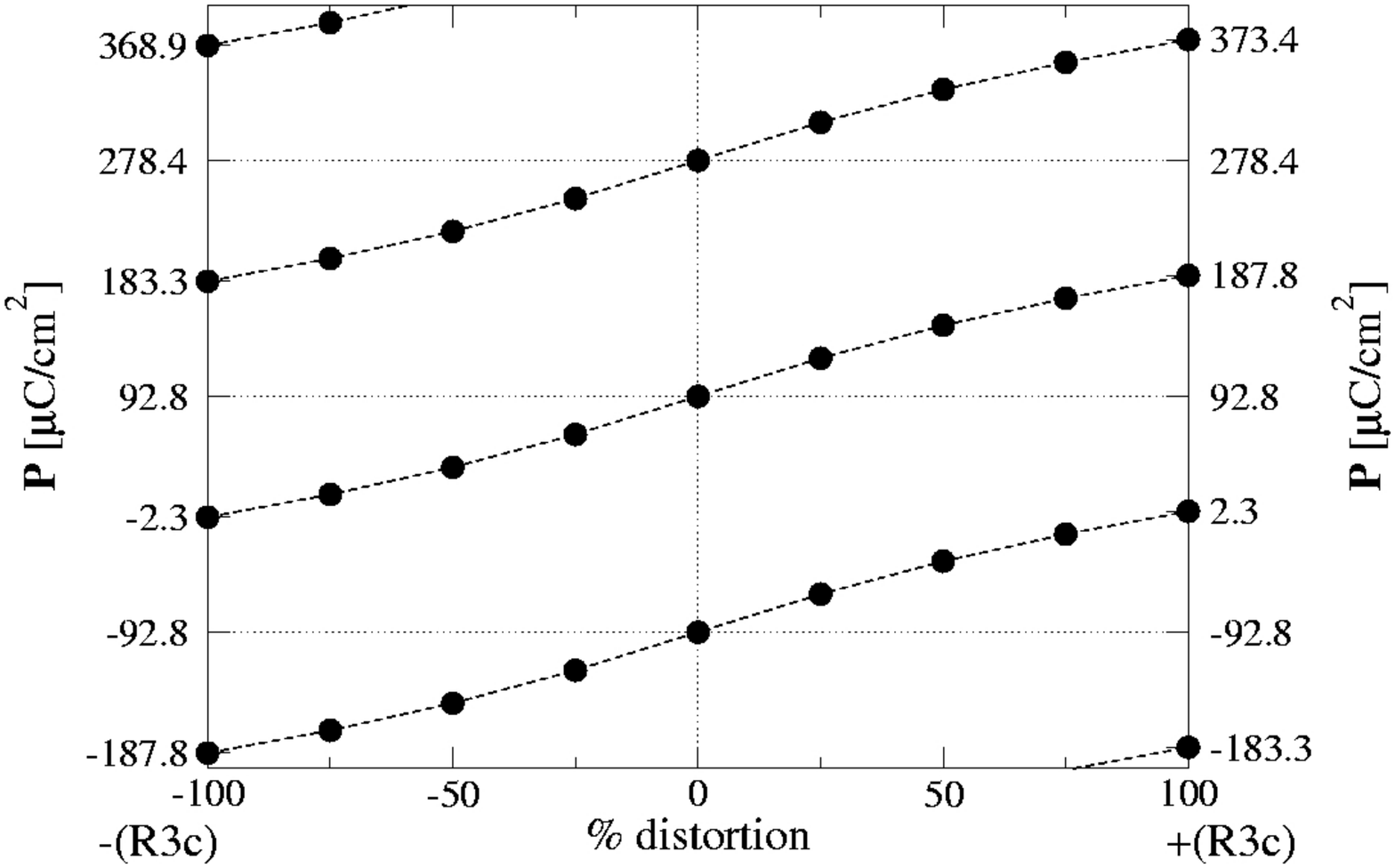}
\caption{Calculated change in polarization, {\bf P}, along a path between the two oppositely polarized -(R3c) and +(R3c) structures ({\bf P} along $[\bar{1}\bar{1}\bar{1}$ and $[111]$) of BiFeO$_3$ through the centrosymmetric cubic structure (0\% distortion). The spontaneous polarization of the R3c structure is the change in polarization between the 0 and 100\% distorted structures, or alternatively half of the change between the -100\% and 100\% distorted structures, along any of the paths, and has the value 95.0 $\mu$C/cm$^2$. The branches are spaced by the polarization quantum, $P_q$ = 185.6 $\mu$C/cm$^2$ along the [111] direction. Note that the polarization lattice of the undistorted centrosymmetric structure is of the half-quantum type, and does not contain zero. From Ref.~\onlinecite{Neaton_et_al:2005}. Copyright (2005) by the American Physical Society.}
\label{BiFeO3}
\end{figure}

Before continuing, we briefly summarize the key aspects of the modern theory of polarization that will be relevant for our discussion of the magnetoelectric multipolization, with an emphasis on the multi-valued polarization lattice and the polarization quantum. No attempt will be made to derive the results here; the reader is directed to the original sources \cite{King-Smith/Vanderbilt:1993,Resta:1993,Vanderbilt/King-Smith:1993,Resta:1994} and tutorial articles \cite{Spaldin:2012}.
The modern theory of polarization tells us that, because of the periodicity of the crystal lattice, the polarization of a periodic solid is a multi-valued quantity consisting of a {\it lattice} of values separated by the polarization quantum, $\vec{P}_q = \frac{e\vec{R}}{V}$. Here $e$ is the electronic charge, $\vec{R}$ is a lattice vector and $V$ is the unit cell volume \cite{Resta:1993,King-Smith/Vanderbilt:1993}. The multi-valuedness arises from the fact that moving an electron by one lattice vector changes the polarization mathematically by $\frac{e\vec{R}}{V}$, but does not change the physical system because of the periodic boundary conditions \cite{Spaldin:2012}. Importantly, the lattice of allowed polarization values must have the same symmetry as the crystal lattice, and so for centrosymmetric materials two types of polarization lattices are possible: $n\vec{P}_q$ (which contains zero) or $(n+ \frac{1}{2})\vec{P}_q$ (which does not contain zero), where $n$ is any positive, negative or zero integer. While apparently inconvenient, the multivaluedness is entirely compatible with experimental reality: Experimentally, only {\it differences} in polarization are measured, and the spontaneous polarization reported for a ferroelectric material is half of the measured polarization difference between two oppositely polarized states. Theoretically, the polarization difference between the two oppositely polarized states (which formally should be connected by an insulating path) must be recorded along the same branch of the lattice of allowed polarization values, and such differences in polarization have the same value whichever branch of the polarization lattice is chosen. An example for the case of multiferroic perovskite-structure bismuth ferrite, BiFeO$_3$ is shown in Fig.~\ref{BiFeO3} \cite{Neaton_et_al:2005}.

Particularly relevant for our discussion is the recent demonstration that the multivaluedness of the polarization yields a unique definition of the surface charge, $\sigma_{\text{surf}}$, associated with the bulk polarization for any chosen surface, given by $\sigma_{\text{surf}} = \vec{P}_{\text{bulk}} \cdot \vec{n}$ \cite{Stengel:2011}. Here $\vec{P}_{\text{bulk}}$ is the dipole moment per unit volume of the bulk unit cell that {\it periodically tiles the semi-infinite solid} containing the surface of interest and $\vec{n}$ is the unit vector normal to the surface. The surface charge is fully determined by the bulk polarization, and different surface charges associated with different choices of surface termination of the same crystallographic plane correspond to different branches of the polarization lattice. 

Like the ferroelectric polarization, the magnetoelectric multipolization contains position in its definition, and so can only be defined in a periodic solid modulo an increment corresponding to displacing a unit of magnetization by a lattice vector. In a system, such as many transition metal oxides, in which the magnetism can be described in terms of local magnetic moments on ions, this multipolization increment can be defined to be the change in multipolization when such a local magnetic moment is displaced by a lattice vector. Note that the situation is more complicated than in the case of the polarization for a number of reasons: First, while the {\it size} of the electron spin is a fundamental quantity just like the electron charge, its {\it orientation} relative to a lattice vector can change. As a result, since the scalar product of a magnetic moment times a lattice vector varies as a magnetic moment is rotated, the multipolization increment also varies, for example when the local magnetic moments are rotated from a centrosymmetric to a non-centrosymmetric arrangement. Second, in the most general case, if there are $l$ magnetic basis atoms in the primitive unit
cell, there can be 9$l$ linearly independent multipolization increments, one for each atom and component of the ${\cal M}$ tensor. This can lead to multiple possibly incommensurate increments existing along certain crystallographic directions. Finally, while the charge on an electron is always $e$, when spin-orbit coupling is taken into account, the magnetic moment associated with an electron differs from $\mu_B$ by the $g$ factor. These factors combined led to the choice of the term {\it increment} rather than quantum to describe the multivaluedness \cite{Ederer/Spaldin:2007}. 

As in the case of the polarization, {\it differences} in magnetoelectric multipolization between related structures are single-valued, and the spontaneous magnetoelectric multipolization can be extracted by recording the change, along the same branch of the multipolization lattice, as the atoms are shifted or the magnetic dipole moments are evolved, from an arrangement that is time-reversal or space-inversion symmetric \cite{Ederer/Spaldin:2007}. Also as in the case of the ferroelectric polarization, the multipolization lattice of a non-magnetoelectric antiferromagnet can be of two types, either containing zero or containing half of a multipolization increment. We explore the implications of this later.

\section{The intrinsic surface magnetization as a bulk property.}

Next, we introduce a definition for the surface magnetization of a semi-infinite slab of an antiferromagnet, as well as a straightforward route to extract it in terms of the bulk magnetoelectric multipolization. Our definition is analogous to the definition of the surface charge in terms of the bulk ferroelectric polarization  \cite{Vanderbilt/King-Smith:1993} and our procedure for extracting it follows the ferroelectric case outlined in Ref.~\cite{Stengel:2011}. We note that, within this definition, the surface magnetization is uniquely defined in terms of the bulk magnetoelectric multipolization, with different chemical terminations of the same crystallographic plane corresponding to different branches of the magnetoelectric multipolization lattice, and that the definition avoids any need to choose a depth for the surface region.  The procedure is as follows: For a particular choice of surface plane orientation and atomic termination, we identify the unit cell that tiles the semi-infinite slab. We then calculate the magnetoelectric multipole of that unit cell, and normalize it to the unit cell volume; this procedure selects the appropriate branch from the multi-valued multipolization lattice of the periodic solid. By analogy with the ferroelectric case we call this $\cal{M}^{\text{bulk}}$, noting that, depending on the symmetry of the system, it can have nine components since it is a 3 by 3 tensor. The surface magnetic dipole per unit area, which we define to the intrinsic surface magnetization, can then be read off directly from the $i, j$ components $\cal{M}^{\text{bulk}}$ tensor, with the first index, $i$ indicating the $x$, $y$ or $z$ orientation of the surface magnetic dipole moments at the surface plane normal to the second index, $j$. We note that this procedure has a rigorous formal basis in the Berry phase theory in the limit of insulating systems in which the electronic wavefunctions can be decomposed into  separate up- and down-spin localized Wannier functions \cite{Thole/Fechner/Spaldin:2016,Stengel:2011}. While not generally the case, it is an excellent approximation in the case of insulating $3d$ transition-metal oxides, with their localized spin magnetic moments, and the extension to other cases, including metallic systems, is intuitively appealing.  We illustrate the procedure in the next section by calculating the intrinsic surface magnetization for the case of Cr$_2$O$_3$.

\begin{figure}
\centering
\includegraphics[scale=0.6]{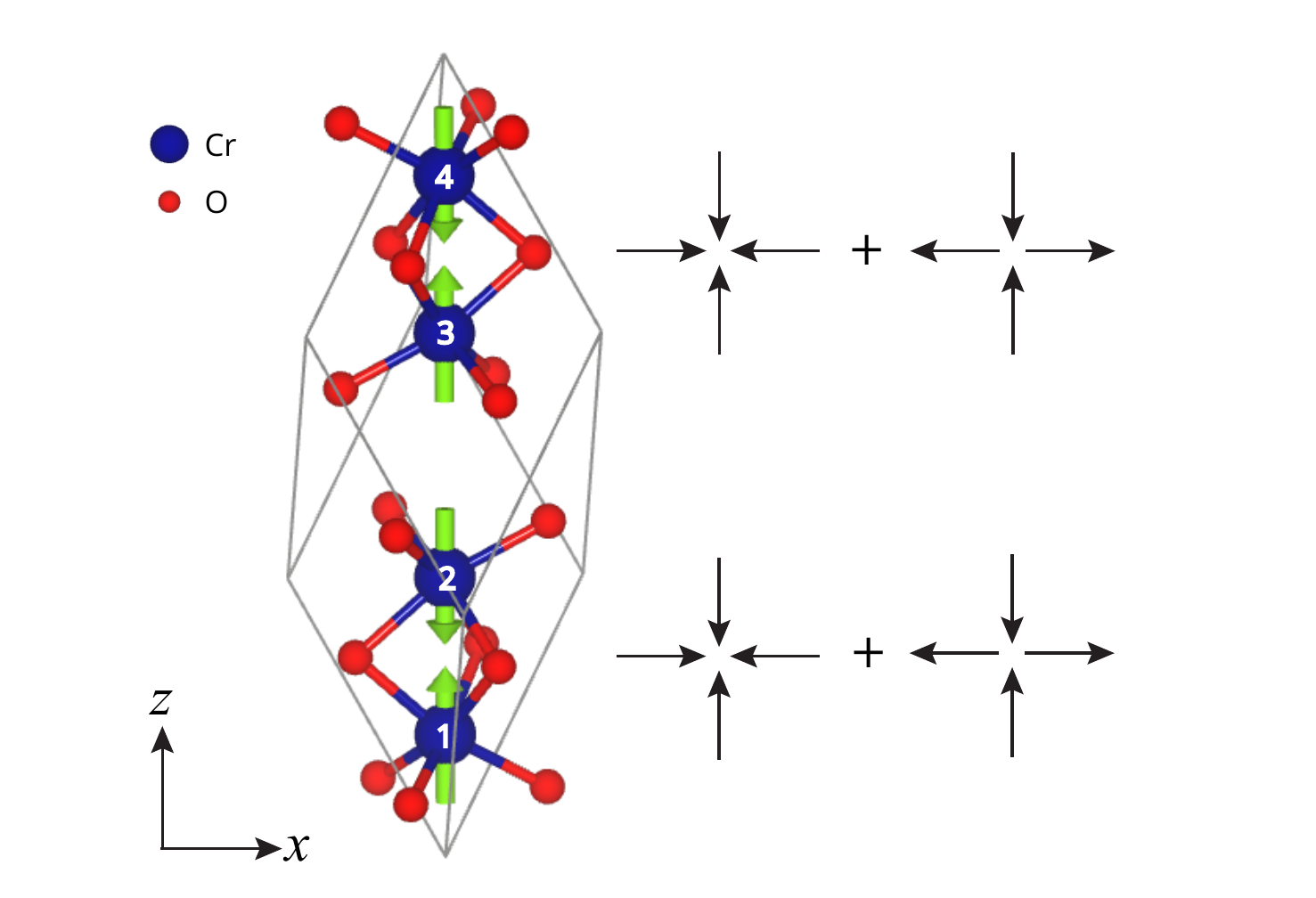}
\caption{Crystal and magnetic structure of Cr$_2$O$_3$. Cr and O ions are shown in blue and red respectively, and the green arrows indicate the directions of the local magnetic moments on the Cr ions (the spins are pointing in the opposite directions). The black lines show the primitive rhombohedral unit cell of the antiferromagnetically ordered structure; the $y$ axis is oriented into the plane. Adapted from  Ref.~\onlinecite{Thole/Keliri/Spaldin:2020}. The patterns of black arrows to the right of the crystal structure show how the magnetic order of Cr$_2$O$_3$ can be decomponsed into the sum of magnetoelectric monopolar and $z^2$ quadrupolar components. For this choice of magnetoelectric domain, both components are of negative sign, that is pointing inwards along $z$.}
\label{Cr2O3fig}
\end{figure}

\subsection{Intrinsic surface magnetization in the prototypical linear magnetoelectric, C{r}$_2$O$_3$}

To celebrate Igor Dzyaloshinskii's 90th birthday, we revisit his prototypical linear magnetoelectric, Cr$_2$O$_3$, in light of this intrinsic surface magnetization concept. Fig.~\ref{Cr2O3fig} shows the primitive rhombohedral unit cell of corundum-structure Cr$_2$O$_3$, below its N\'eel temperature, $T_N$ of $\sim$300K. The formally Cr$^{3+}$ ions are octahedrally coordinated by the oxygen anions, in pairs of distorted octahedra that are face-shared along the vertical $z$ axis and edge-shared with neighboring pairs in the $x-y$ plane. The structural center of inversion is broken by the antiferromagnetic ordering, in which the formally $3 \mu_B$ spin magnetic moments adopt the ``up-down up-down'' pattern along the $z$ axis shown. (Note that the arrows indicate the magnetic moment orientations; the orientation of the spins is of course exactly opposite). The symmetry allows a diagonal linear magnetoelectric effect with different values along the $z$ axis and in the $x-y$ plane; off-axis canting of the spins is symmetry forbidden \cite{Dzyaloshinskii:1960}.

\begin{figure}
\centering
\includegraphics[scale=0.45]{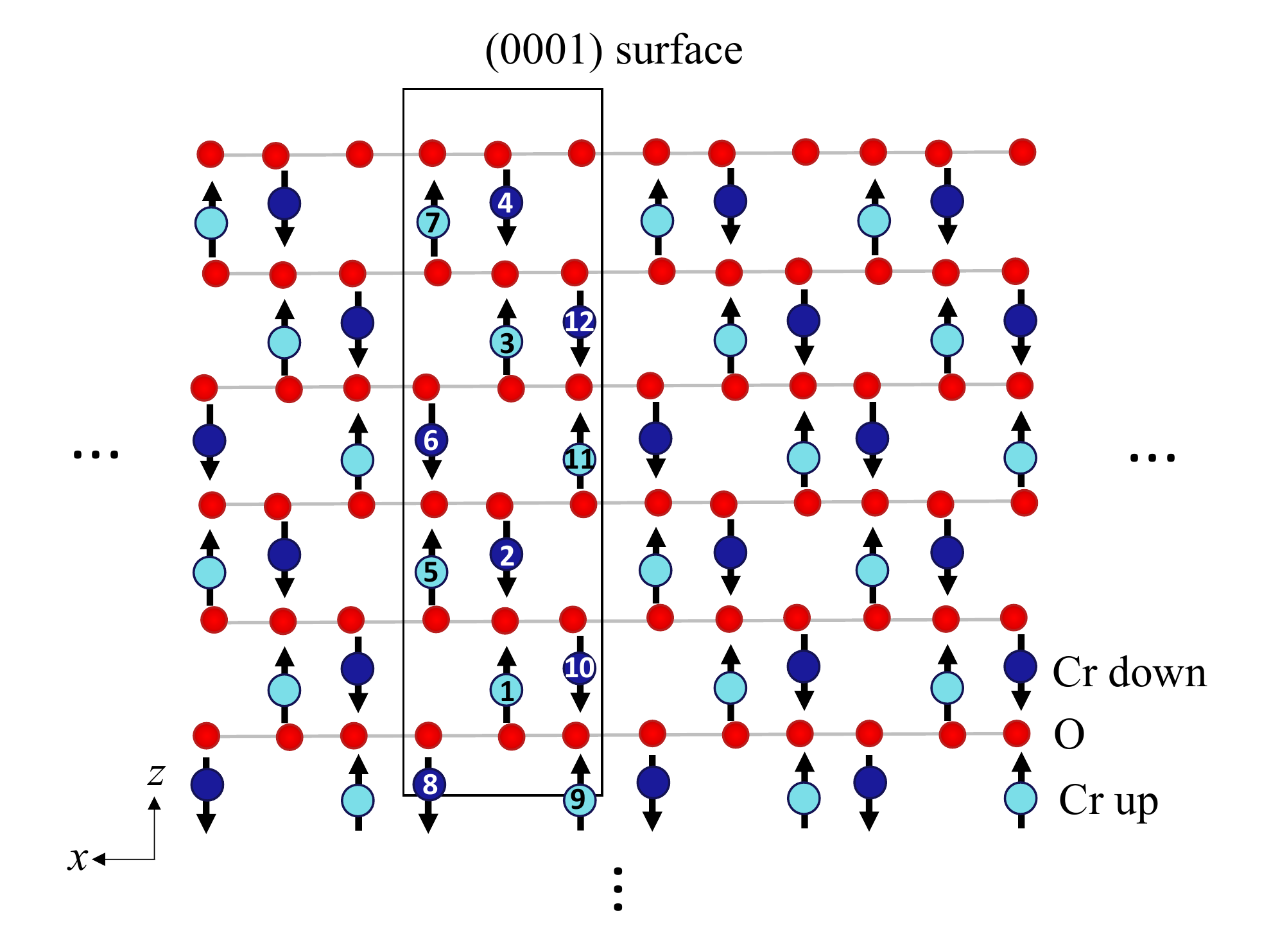}
\caption{A semi-infinite slab of Cr$_2$O$_3$ with a (0001) surface, shown projected down the $y$ axis. Cr and O ions are shown in blue and red respectively, and the arrows indicate the directions of the local magnetic moments on the Cr ions. The black dots, $\cdot \cdot \cdot$, indicate continuation of the structure. The black rectangle shows a choice of hexagonal unit cell,  which, in combination with the numbered Cr ions, can be periodically repeated to tile the slab. The numbers on the Cr ions within the unit cell correspond to those in Table~\ref{Cr2O3Table}.}
\label{Cr2O3_tiling}
\end{figure}

\begin{figure}
\centering
\includegraphics[scale=0.5]{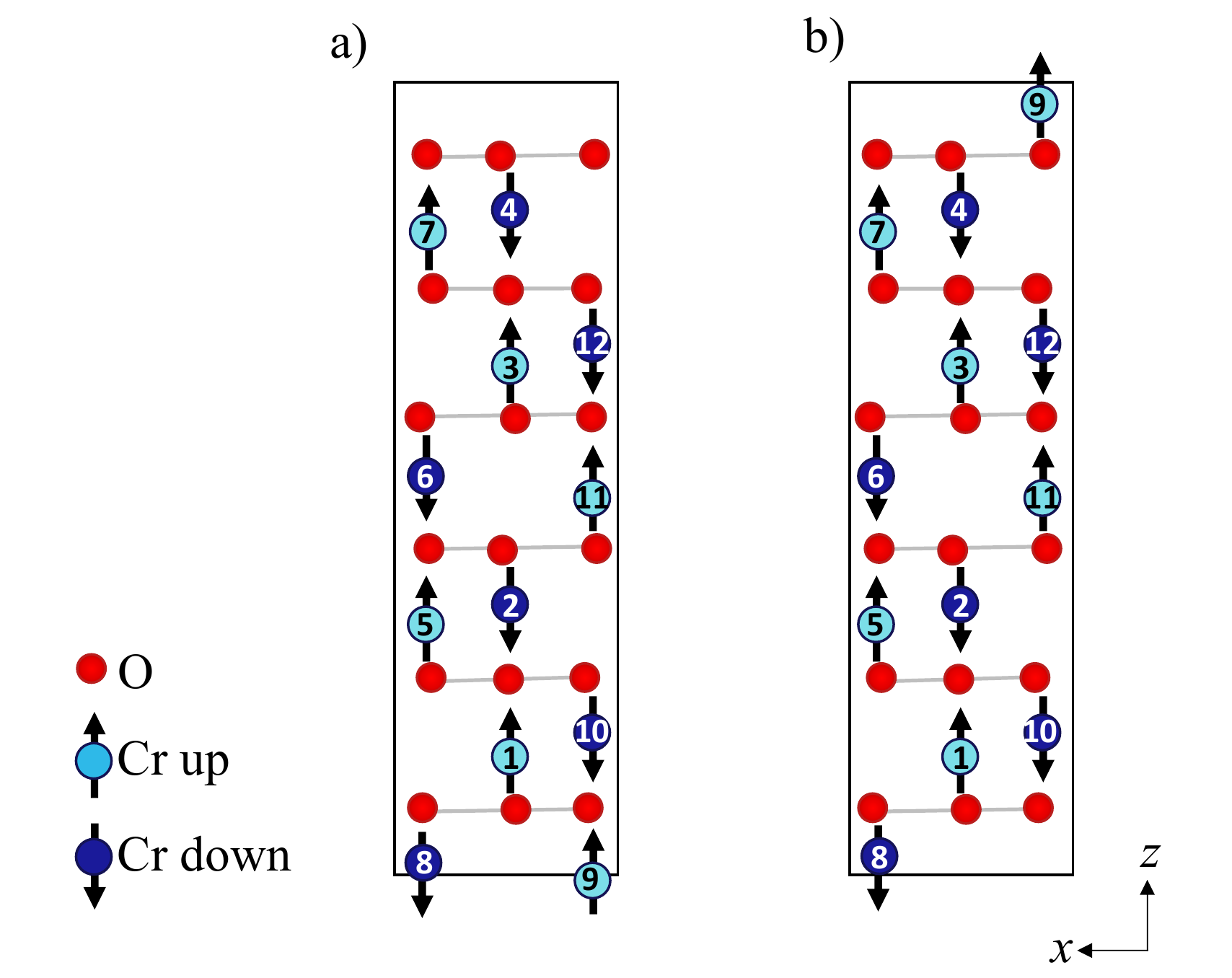}
\caption{Two choices of unit cell (projected down the $y$ axis) that can be used to tile a semi-infinite slab of Cr$_2$O$_3$ with a (0001) surface. a) The surface termination is between an oxygen layer and a Cr-Cr bilayer. The Cr ions have the atomic coordinates given in Table~\ref{Cr2O3Table}. b) The surface splits the Cr-Cr bilayer. The ionic basis in b) differs from that in a) by displacement of one up-moment Cr ion by one lattice vector along the $+z$ direction.}
\label{Cr2O3_unitcells}
\end{figure}

Since we are interested in the (0001) surface, we work with the larger hexagonal cell, containing 12 Cr ions. As illustrated in Fig.~\ref{Cr2O3_tiling}, the hexagonal unit cell, shown with the black rectangle, can be periodically repeated to tile a semi-infinite slab with a (0001) surface, whereas the smaller rhombohedral unit cell can not. The coordinates of the Cr ions, as fractions of the lattice vectors $\vec{R}_1 = (4.91, 0, 0)$, $\vec{R}_2 = (-2.46, 4.25, 0)$,  $\vec{R}_3 = (0, 0, 13.52)$ (in {\AA}) (as calculated in Ref.~\onlinecite{Dehn_et_al:2020}), and their relative magnetic moment orientations $\pm (s = 3 \mu_B)$ are listed in Table~\ref{Cr2O3Table} for the choice of basis shown in Fig.~\ref{Cr2O3_unitcells}a. The unit-cell volume is $V= 282.1$ {\AA}$^3$ and its (0001) surface area is $\frac{282.1}{13.52} = 20.87$ {\AA}$^2$.

\begin{table}
\renewcommand{\arraystretch}{1.4}
\setlength{\tabcolsep}{10pt}
\begin{tabular}{c|ccc|c} \hline \hline
Cr site & $r_i^{\vec{R}_1}$ & $r_i^{\vec{R}_2}$ & $r_i^{\vec{R}_3}$ & $\mu_i^z$ \\ \hline 
4 &  $0$           & $0$          &  $(1-u)$          & $-s$ \\
3 &  $0$           & $0$           & $(\frac{1}{2}+u)$ & $+s$ \\
  
2 &  $0$           & $0$           & $(\frac{1}{2}-u)$ & $-s$ \\
1 &  $0$           & $0$           & $u$             & $+s$ \\ \hline 
  
6 &  $\frac{1}{3}$ & $\frac{1}{6}$ & $(\frac{2}{3}-u)$ & $-s$ \\
5 & $\frac{1}{3}$ & $\frac{1}{6}$ & $(\frac{1}{6}+u)$ & $+s$ \\
 
8 & $\frac{1}{3}$ & $\frac{1}{6}$ & $(\frac{1}{6}-u)$ & $-s$ \\
7 & $\frac{1}{3}$ & $\frac{1}{6}$ & $(\frac{2}{3}+u)$ & $+s$ \\ \hline 
  
12 & $\frac{1}{6}$ & $\frac{1}{3}$ & $(\frac{5}{6}-u)$ & $-s$ \\
11 & $\frac{1}{6}$ & $\frac{1}{3}$ & $(\frac{1}{3}+u)$ & $+s$ \\
  
10 & $\frac{1}{6}$ & $\frac{1}{3}$ & $(\frac{1}{3}-u)$ & $-s$ \\
9 & $\frac{1}{6}$ & $\frac{1}{3}$ & $(u - \frac{1}{6})$ & $+s$ \\ \hline \hline
  \end{tabular}
\caption{Fractional coordinates of the Cr ions in Cr$_2$O$_3$ in the hexagonal setting, with lattice vectors (in {\AA}) $\vec{R}_1 = (4.91, 0, 0)$, $\vec{R}_2 = (-2.46, 4.25, 0)$,  $\vec{R}_3 = (0, 0, 13.52)$ and $u=0.153$, for the unit cell and ionic basis shown in Figs.~\ref{Cr2O3_tiling} and \ref{Cr2O3_unitcells}a. From Ref.~\cite{Dehn_et_al:2020}. The three groupings indicate the three columns of ions along the $\vec{R}_3 = z$ direction that are at different heights relative to the $x-y$ plane, and the Cr site numbers correspond to those in Figs.~\ref{Cr2O3fig}, \ref{Cr2O3_tiling} and \ref{Cr2O3_unitcells}a.  The magnetic moment orientations along the $\vec{R}_3 = z$ direction are indicated in the last column for the antiferromagnetic domain shown in Figs.~\ref{Cr2O3fig},  \ref{Cr2O3_tiling} and \ref{Cr2O3_unitcells}a. }
\label{Cr2O3Table}
\end{table}

We begin by calculating all the non-zero  components of the magnetoelectric multipolization tensor, ${\cal{M}}_{ij}$. The local magnetic moment orientation shown in Figs.~\ref{Cr2O3fig}, \ref{Cr2O3_tiling} and \ref{Cr2O3_unitcells}, and listed in Table~\ref{Cr2O3Table}, in which the moments on neighboring Cr-Cr pairs along the $z$ axis point towards each other, represents one antiferromagnetic magnetoelectric domain. By inspection, we expect this magnetic structure to correspond to a combined  magnetoelectric monopole plus $z^2$ quadrupole both with negative sign, as indicated to the right of the crystal structure in Fig.~\ref{Cr2O3fig}. Approximating the multipole by the sum over the appropriate products of the local moments times their positions, that is using
${\cal{M}}_{ij} = \sum_{\text{ions}} r_i \mu_j$, and normalizing by the unit cell volume, we obtain a multipolization, ${\cal{M}}_{zz}^{\text{bulk}}$ of -2.35 $\mu_B$/nm$^2$, for the atomic positions listed in Table~\ref{Cr2O3Table}. All other components are zero. The opposite antiferromagnetic magnetoelectric domain has the magnetic moments on nearest-neighbor pairs of Cr ions pointing away from each other, and corresponds to a domain with the opposite sign of the magnetoelectric multipole. (We note that Ref.~\onlinecite{Thole/Fechner/Spaldin:2016} compared the Cr$_2$O$_3$ magnetoelectric monopolization obtained from a full density functional calculation using a Berry phase approach, with that obtained by summing over moments of magnitude 3 $\mu_B$ times their positions, and found that the summation method over-estimated the value by around $\frac{1}{3}$.) 
Therefore, by our definition, we find that a slab of Cr$_2$O$_3$ with the domain and surface shown in Fig.~\ref{Cr2O3_tiling}, has a (0001) surface magnetization of -2.4 $\mu_B$/nm$^2$, with the minus sign indicating the downward orientation, that is into the bulk of the material away from the surface. Since all other components of ${\cal{M}}$ are zero, there is no intrinsic magnetization  associated with the surfaces parallel to the [0001] direction, even though Cr$_2$O$_3$ has an in-plane magnetoelectric effect. This apparent conundrum is the result of non-zero monopolar and quadrupolar contributions to ${\cal{M}}_{xx}$ and ${\cal{M}}_{yy}$, which cause the in-plane magneteoelectric response, but which exactly cancel so that ${\cal{M}}_{xx} = {\cal{M}}_{yy} = 0$. We note that, for this domain and surface termination, a bottom surface as shown at the lower edge of the slab in Fig.~\ref{Cr2O3_tiling} would also have its surface magnetization pointing inwards, due to the symmetry of the monopole and the $z^2$ component of the quadrupole. Also, the size and orientation of the surface magnetization do not depend on whether the surface cut is made above or below the oxygen ions, as in the top and bottom surfaces of the figure. An outward pointing surface magnetization can be achieved for this surface type only by reversing the magnetoelectric domain.

 The 2.4 $\mu_B$ value that we obtain for the size of the surface magnetization of the surface shown in Figs.~\ref{Cr2O3_tiling} and \ref{Cr2O3_unitcells}a  is remarkably close to the value of $2.14 \pm 1.5$ $\mu_B$/nm$^2$ reported in Ref.~\onlinecite{Appel_et_al:2019} and the values of 1.6 and 2.3 $\mu_B$/nm$^2$ reported in Ref.~\onlinecite{Woernle_et_al:2020}, all extracted from scanning nitrogen vacancy magnetometry measurements.  The lowest energy (0001) surface in Cr$_2$O$_3$, however, is reported to bisect the Cr-Cr dimer, as shown in Fig.~\ref{Cr2O3_unitcells}b \cite{Freund/Kuhlenbeck/Staemmler:1999}. Comparing with Fig.~\ref{Cr2O3_unitcells}a (which has the value 2.4 $\mu_B$/nm$^2$), we see that this surface is obtained by shifting one up-moment Cr ion (labelled 9 in the figure) by one lattice vector upwards along the $z$ direction. We now repeat the procedure to extract the surface magnetization for the surface termination of Fig.~\ref{Cr2O3_unitcells}b, and obtain the value of +12.0 $\mu_B$/nm$^2$ for the  ${\cal{M}}_{zz}^{\text{bulk}}$ magnetoelectric multipole per unit volume. Note that this differs from the value that we obtained previously by exactly one ${\cal{M}}_{zz}$ multipolization increment, $\frac{|R_3| |s|}{V} = 14.4$ $\mu_B$/nm$^2$, consistent with the displacement of one up magnetic moment (on ion 9) by one lattice vector along $+z$. Therefore, we predict for this surface termination a surface magnetization per unit area of +12.0 $\mu_B$/nm$^2$, pointing away from the surface. This is rather far from the measured value. Note that we assume an ideal stoichiometric surface, neglecting atomic reconstructions and the formation of point defects such as oxygen vacancies, both of which have been reported to occur \cite{Cao_et_al:2015}, as well as, perhaps more importantly, any spin reconstructions. In particular, ion 9 in Fig.~\ref{Cr2O3_unitcells}b lacks the strong nearest- (for example between atoms 4 and 3) and next-nearest- (for example between atoms 4 and 7) neighbor antiferromagnetic exchanges that are primarily responsible for the magnetic ordering in Cr$_2$O$_3$ \cite{Fechner_et_al:2018}. Instead, its largest magnetic interactions are three small and frustrated ferromagnetic third- and fourth-neighbor exchanges (with ions 4 and 7 and their equivalents), and the small antiferromagnetic fifth-neighbor exchange with ion 12 \cite{Fechner_et_al:2018}. It is therefore probable that the moments on ions of type 9 remain disordered and do not contribute to the surface magnetic moment, so that magnetically, the surface of Fig.~\ref{Cr2O3_unitcells}b behaves like that of Fig.~\ref{Cr2O3_unitcells}a, with an additional one-atom thick dead layer. This is particularly likely at the temperature of the scanning nitrogen-vacancy measurements (295K), which is close to the N\'eel temperature \cite{Woernle_et_al:2020}.
Indeed, given that any surface magnetization will cause an external magnetic field \cite{Jiang/West/Zhang:2020} with an associated magnetostatic energy cost at surface edges, steps and domain walls, it will often be favorable for a system to lower reorient its surface magnetic moments, provided that the exchange-energy cost of doing so is not too high. Note that this energy balance would not be detectable in a calculation performed using standard density functional theory, however, since the magnetostatic energy is not taken into account within the density functional formalism, and so would need to be considered separately to compute the relative energies of surfaces with different magnetic terminations from first principles. 
 Note also that the possibility of reducing the surface magnetization by reorienting or disordering the surface magnetic moments is in complete contrast to the situation in ferroelectrics, where charge can only be compensated by the addition or removal of charged species.
 
As an aside, we note that, for the special case of $u=\frac{1}{8}$, the magnetic lattice on the Cr ions in Cr$_2$O$_3$ would have a center of inversion even in its magnetically ordered state, and so we might expect it to not be magnetoelectric. (In practice, the oxygen ions would still break the inversion symmetry). Applying the same analysis described above to extract the ${\cal{M}}_{zz}^{\text{bulk}}$ values in this limit, yields a value of -7.2 $\mu_B$/nm$^2$ for unit cell (a) and +7.2 $\mu_B$/nm$^2$ for unit cell (b) of Fig.~\ref{Cr2O3_unitcells}. Note, first, that both values correspond to half of the multipolization increment of 14.4 $\mu_B$/nm$^2$. Thus, the magnetic lattice of ``fake centrosymmetric'' Cr$_2$O$_3$ is an example of a non-magnetoelectric system with a magnetoelectric multipolization lattice that does not contain zero. For both unit cells, the difference in multipolization between the non-centrosymmetric magnetoelectric structure and the centrosymmetric reference state has the same value, 4.8 $\mu_B$/nm$^2$, which is therefore the {\it spontaneous magnetoelectric multipolization} of Cr$_2$O$_3$.

 \begin{figure*}
\centering
\includegraphics[scale=0.47]{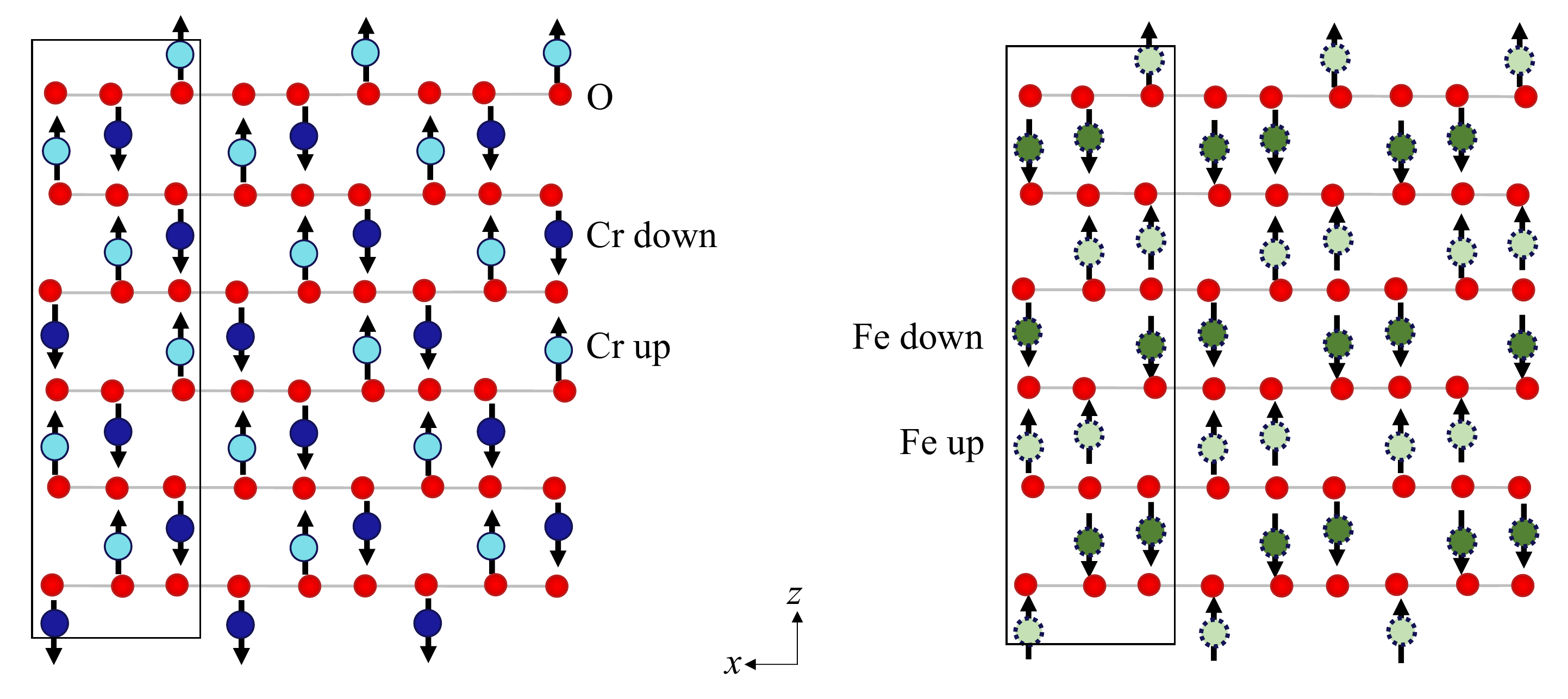}
\caption{Projections of the structures of Cr$_2$O$_3$ (left) and $\alpha$-Fe$_2$O$_3$ (right) down the hexagonal $y$ axis to illustrate the different magnetic orderings. Oxygen ions are shown in red and Cr (Fe) cations in blue (green) with solid (dotted) borders. Light color and up arrows indicate up magnetic moment, and dark color  and down arrows indicate down magnetic moment (below $T_M$ for Fe$_2$O$_3$). The grey horizontal lines are guides to the eye. The upper and lower terminations are the low-energy surface termination for Cr$_2$O$_3$; that for Fe$_2$O$_3$ is unknown. 
\label{Fe2O3_Cr2O3}}
\end{figure*}

Finally for this section, we mention possible experiments for investigating the relevance of the concepts that we have introduced. First, we point out that the scanning nitrogen vacancy technique used in Refs.~\onlinecite{Appel_et_al:2019} and \onlinecite{Woernle_et_al:2020} does not yield simultaneously the orientation of the magnetoelectric domains and the direction of the fringing magnetic field. A measurement of the orientation of the surface magnetic dipole moment for a well-defined single domain of Cr$_2$O$_3$ using an alternative technique would be helpful. Even more desirable would be a direct image of the size and orientation of the magnetic moments in the surface layer, to reveal in particular any deviations from the bulk magnetic order. Here state-of-the-art Lorentz microscopy might be able to provide some information.

\subsection{Intrinsic surface magnetization in non-magnetoelectric, F{e}$_2$O$_3$}

It is interesting to compare the Cr$_2$O$_3$ (0001) surface with that of $\alpha$-Fe$_2$O$_3$, which has the same crystal structure as Cr$_2$O$_3$ but a centrosymmetric ``down-up-up-down'' local magnetic dipole moment structure along the $z$ axis (see Fig.~\ref{Fe2O3_Cr2O3}). Below the N\'eel temperature, $T_N = 955$K, the local magnetic moments lie in the $x-y$ plane with a weak ferromagnetic canting -- indeed this phase of $\alpha$-Fe$_2$O$_3$ was the prototype in which Dzyaloshinskii first proposed the existence of weak ferromagnetism \cite{Dzyaloshinskii:1957} -- and below the Morin temperature, $T_M = 260$K, they are oriented along [0001]. We analyze this latter phase here. In both cases the magnetic ordering does not break inversion symmetry and the linear magnetoelectric effect is not allowed.

Applying the same procedure as for Cr$_2$O$_3$, we first identify that the unit cell shown in Fig.~\ref{Fe2O3_Cr2O3} (right panel) tiles the semi-infinite slab containing the (0001) surface cut bisecting the up-moment bilayer. We then calculate ${\cal{M}}_{zz}^{\text{bulk}}$ for this unit cell and ionic basis and obtain the value of zero. Therefore this surface, and its counterpart obtained by bisecting the down-moment bilayer, have zero surface magnetization. Since the multipolization lattice of Fe$_2$O$_3$ contains zero, intrinsic surface magnetizations equal to integer multiples of the multipolization increment, in this case $-\frac{5}{3} \times 14.4  \approx 24$ $\mu_B$/nm$^2$, due to the $5 \mu_B$ local magnetic moment on the Fe$^{3+}$ ion, are also possible.  It is straightforward to show that a surface plane directly above the down-moment bilayer of this antiferromagnetic domain is tiled by a unit cell that is similar to that shown, but with the uppermost  up-moment ion displaced one unit cell downwards, reducing its multipolization by one increment. As a result it has surface magnetization equal to minus one multipolization increment, that is  $-24$ $\mu_B$/nm$^2$. The surface plane directly above the up-moment bilayer has surface magnetization equal to plus (that is pointing out of the plane) the same value.

 \begin{figure*}
\centering
\includegraphics[scale=0.5]{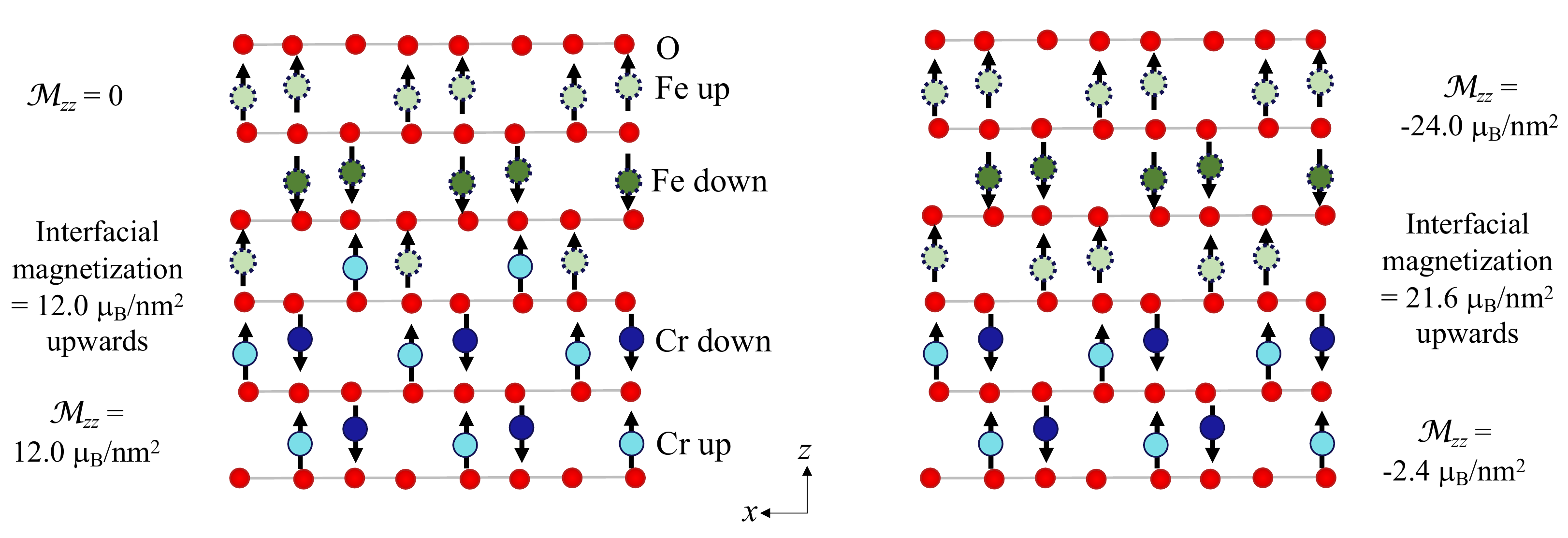}
\caption{Model superlattices of Cr$_2$O$_3$ and Fe$_2$O$_3$ with two possible interfacial configurations. The colors and symbols are as in Fig.~\ref{Fe2O3_Cr2O3}. 
\label{Superlattices}}
\end{figure*}

These results have implications for the magnetism of Cr$_2$O$_3$/Fe$_2$O$_3$ superlattices. While neither the atomistic or magnetic nature of a Cr$_2$O$_3$/Fe$_2$O$_3$ interface is yet known, we see that it is not possible to make such an interface without a net interfacial magnetization, and we can immediately write down the value of the interfacial magnetization -- which is equal to the change in magnetoelectric multipolization across the interface -- for any model case. Two examples are shown in Fig.~\ref{Superlattices}. On the left, the lower Cr$_2$O$_3$ layer has the termination and domain structure that we evaluated earlier to have a multipolization of +12.0 
$\mu_B$/nm$^2$, where the + sign indicates that the corresponding surface magnetization points away from the Cr$_2$O$_3$ surface, that is in the up direction in the Figure. The upper Fe$_2$O$_3$ layer has multipolization zero, so the difference in multipolization, and corresponding interfacial magnetic moment is +12.0 $\mu_B$/nm$^2$, pointing in the upwards direction. On the right, both layers have negative bulk ${\cal{M}}_{zz}$ multipolizations, indicating that their interfacial magnetizations point inwards to the bulk of the layer away from the interface. The multipolization difference between the Fe$_2$O$_3$ and Cr$_2$O$_3$ layers is $(-24.0 - ( -2.4))$ $\mu_B$/nm$^2$ = -21.6 $\mu_B$/nm$^2$, corresponding to an interfacial magnetization of 21.6 $\mu_B$/nm$^2$ in the direction towards the Fe$_2$O$_3$, that is upwards in the figure.

Finally for this section, we mention that the lowest energy termination of a free $\alpha$-Fe$_2$O$_3$ surface is not known.  In spite of tremendous experimental efforts, the preparation of ideal stoichiometric (0001) surfaces of $\alpha$-Fe$_2$O$_3$ has proved prohibitive, and characterization is challenging (for a review see Ref.~\onlinecite{Parkinson:2016}). The reported behavior depends strongly on annealing conditions, often with massive surface restructuring and combinations of Fe$_3$O$_4$ and Fe$_{1-x}$O contributions to the surface layers \cite{Schoettner_et_al:2019}. To our knowledge, any role of surface magnetization associated with the multipolization increment in destablilizing the (0001) surface has not been discussed. It is also intriguing to consider whether surface magnetization could play a role in the formation of the recently reported complex topological domain structure \cite{Chmiel_et_al:2018}.

\subsection{Intrinsic surface magnetization in non-magnetoelectric materials with half-increment-containing multipolization lattices}

It is tempting to assume that non-magnetoelectric materials are immune from any surface magnetization resulting from the truncation of their bulk magnetoelectric multipolization. We have already seen, however, that for the case of non-magnetoelectric Fe$_2$O$_3$, even though zero is an allowed value for its magnetoelectric multipolization, the multivaluedness leads to non-zero surface magnetizations for certain terminations. The situation is further complicated for non-magnetoelectric materials whose multipolization lattice contains half a multipolization increment, since in this case surface planes exist for which zero is not an allowed value for any stoichiometric termination. We discuss this scenario next for the case of a model one-dimensional chain.

\begin{figure}
\centering
\includegraphics[scale=0.2]{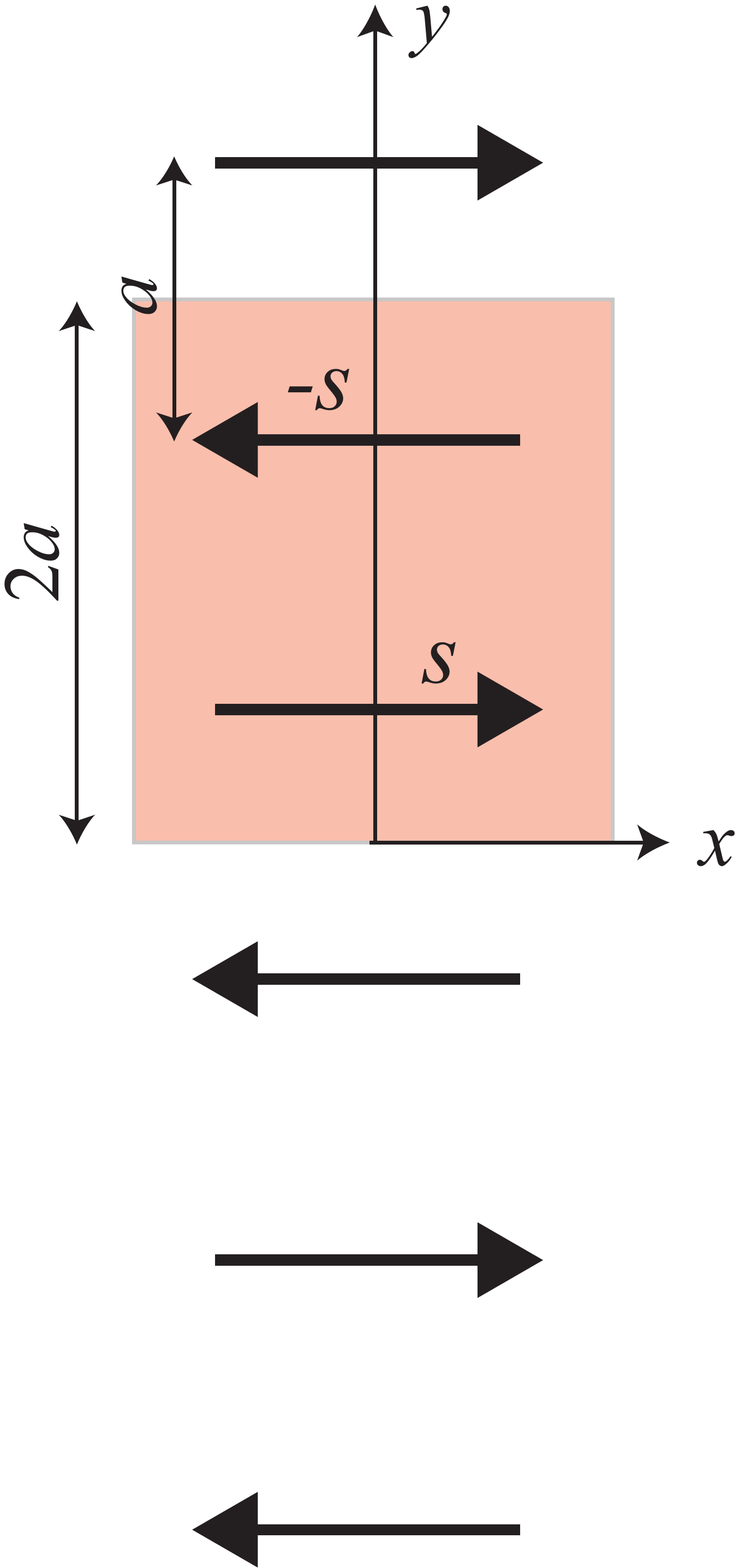}
\caption{A model 1D centrosymmetric and therefore non-magnetoelectric system in which local magnetic dipole moments, $s$, oriented antiferromagnetically along the $x$ axis, are equally spaced by $a$ along $y$. One choice of unit cell is shown in pink. Adapted from Ref.~\onlinecite{Ederer/Spaldin:2007}. }
\label{t_1Dchain}
\end{figure}

As a straightforward example of a non-magnetoelectric system that does not contain zero in its multipolization lattice we take the example of one-dimensional chain of alternating localized magnetic dipole moments, as shown in Fig.~\ref{t_1Dchain} 
\cite{Ederer/Spaldin:2007}.
The moments, with magnitude $s = 1 \mu_B$ say, are spaced a
distance $a$ apart from each other along the $y$ axis, and are
alternating in orientation along $\pm x$ with two oppositely oriented magnetic moments in each
unit cell. 
It is straightforward to see that the arrangement of magnetic moments in the lattice is
space-inversion symmetric with respect to each moment site and so there can be no 
macroscopic magnetoelectric multipolization. In addition, while the magnetic ordering breaks time-reversal symmetry at the unit-cell level, time inversion combined with a translation of all moments by distance $a$ along the $y$ direction is a symmetry operation, and so time-reversal
symmetry is not broken \emph{macroscopically}, again prohibiting a macroscopic multipolization.

The single unit cell highlighted in Fig.~\ref{t_1Dchain}, however, has a non-zero and non-trivial $yx$ component of its magnetoelectric multipolization tensor, ${\cal{M}}_{yx}^{\text{bulk}} = \sum_i y_i \mu^x_i$. Here the sum is over the $i$ ions in the unit cell, $y_i$ indicates their position along the $y$ direction and $\mu^x_i$ the $x$ component of their magnetic moment. Performing the summation and normalizing to the unit cell ``volume'', which for the one-dimensional unit cell is just its length $2a$, one obtains  ${\cal{M}}_{yx}^{\text{bulk}}= -\frac{s}{2}$. (This is composed of a toroidization contribution, $-{\vec{T}_z} = -\frac{s}{4} \hat{z}$ and a quadrupolization, $Q_{yx} = -\frac{s}{4}$.) The magnitude of ${\cal{M}}_{yx}^{\text{bulk}}$ is half of the corresponding multipolization increment, which is the change in multipolization when a magnetic moment is displaced by a lattice vector, in this case $\frac{s \times 2a}{2a}  = s$. Therefore, in spite of the fact that the system is not magnetoelectric, its multipolization lattice does not contain zero, but instead is of the non-trivial type that contains the half multipolization increment. Performing the calculation of ${\cal{M}}_{yx}^{\text{bulk}}$ for other choices of unit cell or basis would yield other values on the multipolization lattice, separated from this value by $n s$, but would never yield zero. The non-zero bulk multipolization results in a surface magnetic dipole moment of  ${\cal{M}}_{xy}^{\text{bulk}} = -\frac{s}{2}$ per unit area, pointing along $-x$, when the system is truncated above the highlighted unit cell shown; truncating half a unit cell lower, above the layer of magnetic moments pointing along $+x$, would result in a surface magnetic dipole moment of the same magnitude but pointing along $+x$.

Finally, we discuss briefly an analogous ferroelectric case: the intrinsic surface charge in the non-polar centrosymmetric III-III ideal cubic perovskite structure LaAlO$_3$. It is trivial to show, from summing the formal charges La$^{3+}$, Al$^{3+}$ and O$^{2-}$ multiplied by their positions in the unit cell, that the polarization lattice of LaAlO$_3$ is of the form $\frac{\vec{P}_q}{2} + n \vec{P}_q$ and does not contain zero. As a result, its \{100\} surfaces have a bound charge \cite{Stengel:2011}. In contrast, the II-IV perovskite SrTiO$_3$ has a zero-containing polarization lattice. 
A [001]-oriented heterostructure composed of the two materials has a polar discontinuity at its interface of magnitude exactly $\frac{1}{2}P_q^{[001]}$, where $P_q^{[001]}$ is the component of the polarization quantum in the [001] direction. To ensure electrostatic stability, this discontinuity must be screened \cite{Nakagawa/Hwang/Muller:2006}, with the sign of the required screening charges determined by the nature of the interface (positive for AlO$_2$ / SrO and negative for TiO$_2$ / LaO) \cite{Stengel:2011}. In the latter case the accumulated electrons are mobile carriers in the bottom of the SrTiO$_3$ valence band, leading to a conducting interface that is even superconducting at low temperatures \cite{Reyren_et_al:2007}.

\section{Summary and Outlook}

In summary, we have reviewed the phenomenology of magnetoelectric multipolization in bulk, periodic solids, and provided an analogy with various aspects of the ferroelectric polarization. 
We showed that the analogy provides a particularly convenient picture of the surface magnetization that is associated with magnetoelectric materials \cite{Belashchenko:2010, Andreev:1996}, and we provided the following straightforward recipe to extract it from the bulk magnetoelectric multipolization for a given surface plane:
\begin{description}
\item[1] For the surface plane and chemistry of interest, identify the unit cell and ionic basis that tiles a semi-infinite slab of the system.
\item[2] Calculate the components of the bulk magnetoelectric multipolization, ${\cal M}_{ij}^{\text{bulk}}$, using this unit cell and basis of ions, and normalizing it to the unit cell volume.
\item[3] The non-zero components of ${\cal M}_{ij}^{\text{bulk}}$ that have a contribution normal to the surface plane then give directly the size and orientation of the intrinsic surface magnetization.
\end{description} 
We argued that such an intrinsic surface magnetization is possible even at the surface or interface of a non-magnetoelectric material, and distinguished two cases: In non-magnetoelectric materials whose multipolization lattice contains zero it is always possible to choose a stoichiometric termination with zero magnetic moment for any choice of surface plane, although this might not necessarily be the lowest energy termination. In non-magnetoelectric materials whose multipolization lattice contains the half-multipolization increment, in contrast, surface planes exist for which an intrinsic magnetic moment can not be avoided for stoichiometric terminations. 

We mentioned some phenomena for which these concepts might be relevant and which could provide interesting directions for future work. In particular, the intrinsic surface magnetization arising from the magnetoelectric multipolization could have implications for the relative stability of antiferromagnetic surfaces and interfaces, the formation of antiferromagnetic domains, and the mechanism of exchange-bias coupling. A comparative survey of the exchange biasing behavior of magnetoelectric versus non-magnetoelectric antiferromagnets, as well as non-magnetoelectric antiferromagnets with zero- and half-increment-containing multipolization lattices could be fruitful. Finally, we suggested some experiments that could be used to verify or disprove our proposals, and we hope, in the spirit of Igor Dzyaloshinskii, that this manuscript motivates future experimental work in these directions.

\section{Acknowledgments} 

This work was supported by the K\"orber Foundation, the European Research Council (ERC) under the European Union’s Horizon 2020 research and innovation programme grant agreement No 810451 and by the ETH Zurich. Thanks to Igor Dzyaloshinskii for the many inspiring discussions and fruitful collaborations, and to Sayantika Bhowal, Christian Degen, Manfred Fiebig, Pietro Gambardella, Kane Shenton, Tara To\v{s}i\'c and Xanthe Verbeek for helpful comments on the manuscript.

\newpage
\bibliography{Nicola.bib}

\end{document}